\documentclass[apjl]{emulateapj}
\usepackage[latin2]{inputenc}
\usepackage{amsmath}
\usepackage{graphicx}
\usepackage{url}

\newcommand{\ff}{{\tt FAST-FORWARD} }

\begin{document}
\title{Milky Way mass and potential recovery using tidal streams in a realistic halo}
\author{Ana Bonaca\altaffilmark{1}, Marla Geha\altaffilmark{1}, Andreas H. W. K\" upper\altaffilmark{2,5}, J\" urg Diemand\altaffilmark{3}, Kathryn V. Johnston\altaffilmark{2}, David W. Hogg\altaffilmark{4}} 
\altaffiltext{1}{Department of Astronomy, Yale University, New Haven, CT 06511; {ana.bonaca@yale.edu}}
\altaffiltext{2}{Department of Astronomy, Columbia University, New York, NY 027}
\altaffiltext{3}{Institute for Computational Sciences, University of Z\" urich, 8057 Z\" urich, Switzerland}
\altaffiltext{4}{Center for Cosmology and Particle Physics, Department of Physics, New York University, 4 Washington Place \#424, New York, NY 10003}
\altaffiltext{5}{Hubble Fellow}

\begin{abstract}
We present a new method for determining the Galactic gravitational potential based on forward modeling of tidal stellar streams.
We use this method to test the performance of smooth and static analytic potentials in representing realistic dark matter halos, which have substructure and are continually evolving by accretion.
Our {\tt FAST-FORWARD} method uses a Markov Chain Monte Carlo algorithm to compare, in 6D phase space, an ``observed'' stream to models created in trial analytic potentials.
We analyze a large sample of streams evolved in the Via Lactea II (VL2) simulation, which represents a realistic Galactic halo potential.
The recovered potential parameters are in agreement with the best fit to the global, present-day VL2 potential.
However, merely assuming an analytic potential limits the dark matter halo mass measurement to an accuracy of 5 to 20\%, depending on the choice of analytic parametrization.
Collectively, mass estimates using streams from our sample reach this fundamental limit, but individually they can be highly biased.
Individual streams can both under- and overestimate the mass, and the bias is progressively worse for those with smaller perigalacticons, motivating the search for tidal streams at galactocentric distances larger than 70\;kpc.
We estimate that the assumption of a static and smooth dark matter potential in modeling of the GD-1 and Pal5-like streams introduces an error of up to 50\% in the Milky Way mass estimates.
\end{abstract}
\maketitle

\section{Introduction}
The mass density profile, shape and extent of dark matter halos that surround galaxies provide important clues to how galaxies form, and test the properties of the dark sector and the fundamental cosmological model. Given the vast stellar datasets and unique perspective on our own Galaxy, we might expect the mass distribution for the Milky Way to be well understood. However, estimates for the overall mass of the Galaxy are currently uncertain by a factor of two \citep[see Figure~7 in][and references therein]{barber2013}. 

A major hurdle in measuring the Milky Way mass has been the scarcity of tracers in the outer halo. So far, the best measurements were obtained from satellite galaxy \citep[e.g.,][]{watkins2010} and halo star kinematics \citep[e.g.,][]{deason2012}. These methods usually implicitly assume that the stellar halo is a relaxed structure with objects on randomized orbits.

However, extended substructures discovered in the Galactic halo reveal that the Milky Way halo is neither dynamically old nor relaxed \citep[e.g.,][]{majewski2003,belokurov2006,juric2008,bonaca2012}. The presence of tidal debris invalidates gravitational potential recovery methods based on the assumption of random orbits, as neither orbits nor orbital phases are equally represented in an unrelaxed halo. This has been shown through simulations of Milky Way-like halos, which suggest that even large samples of hundreds of outer halo stars may actually be on only a handful of independent orbits \citep{bj2005}. Assuming that orbits in a truly unrelaxed halo are randomized introduces a bias of $\sim20\%$ in mass estimates \citep{yencho2006}. 

Like satellite galaxies and halo stars, tidal streams can be used as tracers of the gravitational potential. In particular, their extent over tens of degrees on the sky offers a unique glimpse into the Galactic halo shape. The largest identified tidal debris in the Galactic halo is the currently disrupting Sagittarius dwarf spheroidal \citep[Sgr;][]{ibata2001b}. The Sgr stream is an example of a dynamically hot stream, due to its large width in both position \citep[$\gtrsim20^\circ$;][]{belokurov2006} and velocity space \citep[$\gtrsim20$\;km/s;][]{koposov2013}. It is also very long; its tails have been traced to wrap more than 360 degrees using M giants, blue horizontal branch and turn-off stars \citep{majewski2003, belokurov2014, piladiez2013}. Despite this wealth of observational data, there is little agreement on the halo shape from Sgr models \citep{helmi2004,johnston2005,law2010,deg2013,veraciro2013,ibata2013,gibbons2014}. Most significantly, uncertainties in the amount of progenitor rotation \citep{penarrubia2010}, the presence of bifurcations in both leading and trailing arms \citep{koposov2012b} and orbit evolution due to dynamical friction \citep{ibata2001c} affect the recovery of the Galactic potential using Sgr.

Cold streams might be expected to provide more stringent constraints on the potential as compared to the Sgr stream \citep{lux2013}. These streams originate from smaller stellar systems such as lower mass dwarf galaxies and globular clusters \citep[e.g.,][]{odenkirchen2003,gd,belokurov2006,grillmair2009,bonaca2012,koposov2014,martin2014,bernard2014}. Both classes of stellar systems are dispersion dominated, so their disruption mechanism is better understood and the potential recovery is less affected by the assumptions on the progenitor's properties. The warmer Orphan stream yielded estimates of the total mass within 60\;kpc \citep{newberg2010,sesar2013}, while \citet{koposov2010} used the 6D phase space information from the colder GD-1 stream to constrain both the halo mass and shape inside the Solar circle. 

Irrespective of their dynamical temperature, tidal streams have so far been modeled exclusively in parametric potential forms which are smooth and static. Yet, $N$-body simulations of structure formation show that dark matter halos are clumpy and evolving. For example, Milky Way-mass halos accrete on the order of $\sim$10\% of their present mass in the last 6\;Gyr \citep{diemand2007}. Furthermore, $\sim$10\% of their total halo mass is in subhalos \citep[e.g.,][]{diemand2008}, which can influence the progenitor orbit \citep[e.g.,][]{veraciro2013}, heat up streams upon close encounters \citep[e.g.,][]{yoon2011}, or create density inhomogeneities \citep[e.g.,][]{sgv2008}.

In this paper, we test biases introduced into recovering the gravitational potential when potential complexity is not taken into account. To do this, we compare two sets of ``test'' streams, the first is formed by inserting cold stellar streams into a realistic, high-resolution dark matter-only simulation, and the second is formed through evolution in a simplified, smooth and static potential. We recover the gravitational potential using these sets of streams.

The structure of the paper is as follows. We start by describing the Via Lactea II (VL2) simulation that we use as a realistic, Milky Way-like dark matter halo in \S\ref{sec:simulation}. We test how well commonly used profiles represent the VL2 potential, and provide triaxial potential fits (\S\ref{sec:fits}). We also explore how the recovered best-fit parameters depend on the radial extent of the data, and how they change over time as the potential evolves (\S\ref{sec:evolution}). Next, we describe a large sample of stellar streams created in the VL2 (\S\ref{sec:vl2streams}) and an analogous sample of streams created in analytic potentials (\S\ref{sec:control}). Analytic potential parameters were obtained by fitting the VL2 potential field. We present a novel method for forward modeling of stellar streams, the \ff method, in \S\ref{sec:modeling}. The results of potential recovery on both stream samples are presented in section \S\ref{sec:results}, while the differences between the samples are discussed in \S\ref{sec:discussion}. We also put our results in the context of the known Milky Way streams in \S\ref{sec:mw}, and conclude in \S\ref{sec:summary}.

\section{Testbed for the Galactic potential}
\label{sec:simulation}
High-resolution simulations are essential for interpreting Galactic studies in a cosmological context. We use the Via Lactea II to represent the Galactic potential, in which tidal streams are formed. Here we provide a brief overview of the simulation in \S\ref{sec:vl2}, and analytic fits to its potential in \S\ref{sec:fits} and \S\ref{sec:evolution}. Stream formation is described in the next section \S\,\ref{sec:streams}.

\subsection{Via Lactea II simulation}
\label{sec:vl2}
Via Lactea II (VL2) is a high resolution dark matter simulation of a Milky Way-like halo \citep{diemand2008}. The simulation contains 1.1 billion particles, each of mass $4.1\times10^3\;\rm M_\odot$, evolved using the PKDGRAV2 tree-code \citep{stadel2001} and a force resolution of 40~pc. The simulation was run from redshift $z\simeq100$ to the present, with adaptive time steps set to 1/16 of the local dynamical time. The main halo experienced no recent major merging events, but grew through minor mergers for the past 6~Gyr. The final mass within the radius enclosing 200 times the mean matter density, $r_{200}=402\;\rm kpc$, is $M_{200}=1.9\times10^{12}\;\rm M_\odot$. This falls in the range of most Milky Way mass determinations \citep{barber2013}. At a distance of $\sim800\;\rm kpc$ from the main VL2 halo, there is another massive halo \citep[$6\times10^{11}\;\rm M_\odot$;][]{teyssier2012}, thus VL2 is similar to the Milky Way not only in terms of its mass and accretion history, but also as a halo in a Local Group-like environment.

Dark matter halos have a universal spherically averaged density profile, regardless of their mass or cosmological parameters \citep[NFW;][]{nfw}. The density profile of VL2 is also well described by a cusped, NFW-like profile \citep{diemand2008}. Lifting the requirement of spherical symmetry in modeling of dark matter halos, \citet{jing2002} showed that they are, in general, triaxial. In VL2, \citet{zemp2009} detected triaxiality in halo shape from deviations of local density from the spherically averaged values. For ease of calculations, most potential recovery methods introduce triaxiality in the potential, instead of density. In the remainder of this section, we provide fits of the triaxial NFW and logarithmic potentials to the Via Lactea II simulation (\S\,\ref{sec:fits}). Furthermore, we show how they have evolved during the last 6 Gyrs (\S\,\ref{sec:evolution}).

\begin{figure*}
\begin{center}
\includegraphics[scale=0.7]{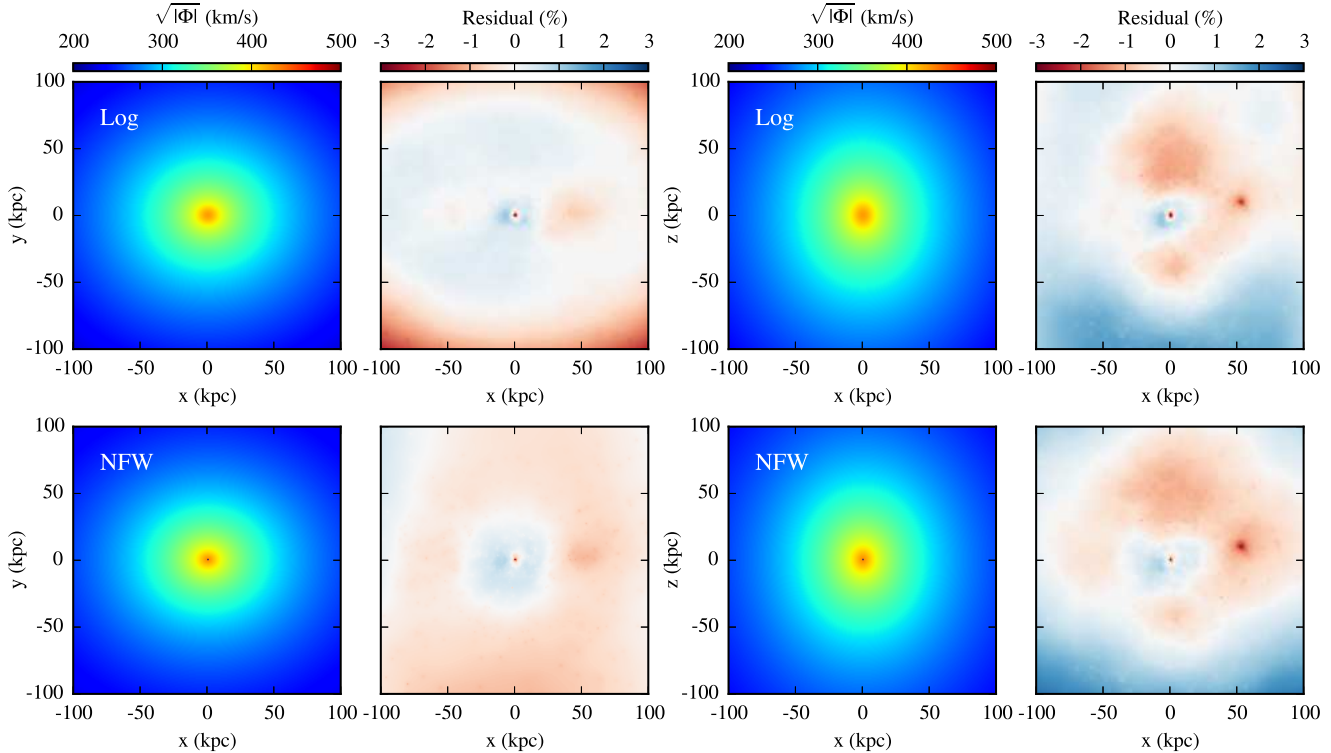}
\caption{Triaxial parametric fits to the present-day Via Lactea II potential, logarithmic in the top row, and NFW on the bottom. The best fit to the $z=0$ plane is shown in the first column, color-coded by $\sqrt{|\Phi|}$ (km/s), with the residuals normalized to the exact VL2 potential in the second column. Similarly, the third column shows the best fit to the $y=0$ plane, and the fourth column has the corresponding residuals. Although there are inhomogeneities in the residuals, they are on the order of $\lesssim3\%$. 
}
\label{fitpot}
\end{center}
\end{figure*}

\subsection{Parametric fits to the VL2 potential}
\label{sec:fits}
The true form of dark matter halo profiles is unknown. Historically, Navarro-Frenk-White (NFW) and logarithmic profiles have been the analytic potential forms most commonly used to represent dark matter halos. Logarithmic halos are motivated by observed flat-rotation curves of galaxies \citep{rubin1980}, while the NFW profile provides a good fit to the spherically averaged profiles of simulated dark matter halos \citep{nfw}. In this section, we assess performance of both of these potential forms on the Via Lactea II potential.

We studied the VL2 potential on a random sample of 100,000 simulation particles extracted at redshift $z=0$. The exact gravitational potential at each particle position was calculated by summing up the contributions from all the other particles, treating each as a Plummer sphere with a characteristic scale of $50\;\rm pc$. Next, the VL2 potential was modeled with these commonly used parametrizations:
\begin{itemize}
 \item a logarithmic potential of form
 \begin{equation}
 \Phi_{log}=V_c^2 \ln (r^2+R_c^2) ,
 \label{log}
 \end{equation}
 where $V_c$ is the circular velocity and $R_c$ is the core radius
 
 \item a NFW potential of form
 \begin{equation}
 \Phi_{NFW}=-V_h^2 (R_h/r) \ln (1+r/R_h) ,
 \label{nfw}
 \end{equation}
 where $R_h$ is the scale radius, and $V_h$ is the circular velocity at $R_h$.
\end{itemize}
In both potential forms, the radial distance $r$ takes into account halo triaxiality:
\begin{equation}
 r^2=C_1x^2+C_2y^2+C_3xy+(z/q_z)^2 .
 \label{rtriax}
\end{equation}
We follow \citet{law2010} in defining the constants $C_i$:
\begin{align*}
 C_1 &= \cos^2 \phi/q_1^2 +\sin^2\phi , \\
 C_2 &= \cos^2 \phi +\sin^2\phi/q_1^2 , \\
 C_3 &= 2\sin\phi\cos\phi(1/q_1^2-1) ,
\end{align*}
where $q_1$ is the ratio between the $x$ and $y$ axes, $q_z$ is the ratio between the $z$ and $y$ axes, and $\phi$ is the rotation angle around the $z$ axis. The chosen coordinate system is aligned with one of the dark matter halo symmetry planes. A general orbit in these triaxial potentials does not necessarily have conserved integrals of motion \citep[e.g., actions][]{bt}. In that sense, they are not integrable, and the action-angle based potential recovery methods are not guaranteed to always work with these potential forms.

Both potential forms have five free parameters: two defining typical circular velocity and scale radius of the halo, and three for its shape and orientation. The best-fit parameters were found by a least-square minimization of the difference between the analytical model and the exact potential at the positions of the simulation particles. The minimization was performed 500 times on a bootstrapped sample of simulation particles, thus providing both the best-fit parameters and their uncertainties. The best-fit analytic models to the inner $100\;\mathrm{kpc}$ are shown in Figure~\ref{fitpot}, with the logarithmic potential on the top and the NFW profile on the bottom. The first column shows the potential in the $x-y$ plane, color-coded by a square root of the gravitational potential, $\sqrt{|\Phi|}$. The second column has normalized, $\rm (model-VL2)/VL2$, potential residuals in the same plane. Similarly, the potential in the $x-z$ plane is shown in the third column, with the corresponding residuals in the fourth. Both the logarithmic and NFW forms are very good approximations to the exact VL2 potential on the global scale; the deviation of the logarithmic best-fit model from the exact potential is on the order of 3\%, while the NFW fit is even better with residuals smaller than 1\%.

Although the residuals are small, they show a radial structure, indicating that some of the potential parameters have radial dependence. In Fig~\ref{potevo} we investigate this dependence by probing progressively larger VL2 radii ($20\;\rm kpc$~$<r_{\rm max}<150\; \rm kpc$) when fitting model potentials. This encompasses the radial range probed by our simulated streams (for more details, see \S\ref{sec:streams}). As before, we start by calculating the potential at each particle due to all other particles in the simulation. We then select particles within a maximum radius $r_{\rm max}$, and fit the potential form to just those particles. The radial dependence of the best-fit parameters, including their $1\sigma$ uncertainties, is shown in Figure~\ref{potevo} (black points) for the logarithmic on the left and the NFW potential on the right. Values of all potential parameters change with radius on the order of $\lesssim10\%$. The only exception is the scale radius for a logarithmic fit, which varies 40\%. The rate of change is similar for both potential forms used. Furthermore, the shape parameters have very similar values in the logarithmic and NFW best-fit models. The halo profile is mildly triaxial, with $x$ axis being the shortest, and $z$ the longest. The axis ratios $q_i$ have the extremes at $40-60$~kpc, beyond which $q_1$ increases and $q_z$ decreases with radius, making the halo more spherical on average. The rotation of the $x-y$ plane around the $z$ axis is $\phi\simeq90^\circ$, i.e., the coordinate system is roughly aligned with the halo axes. The variations in recovered potential parameters presented in Figure~\ref{potevo} indicate that even if an exact measurement of the potential is available, fitting a parametric potential form will yield different results depending on the radial extent of the available data.

The first goal of gravitational potential recovery is to determine a halo's mass and shape. Once the potential form is known, the density is easily obtained using Poisson's equation. Integrating the density field out to a spherical radius, we can obtain an estimate of the enclosed mass. This estimate might be different than that obtained when fitting the density field directly. However, the transition from potential to density is often made in Galactic potential studies, and here we assess its accuracy. The VL2 halo mass inside 150\;kpc estimated using the best-fit logarithmic potential is $M_{\rm tot,log}=1.37\times10^{12}\;{\rm M}_\odot$, and for the NFW potential equals $M_{\rm tot,NFW}=1.09\times10^{12}\;{\rm M}_\odot$. A direct measure of mass in the inner 150\;kpc of the VL2 yields $M_{\rm tot,VL2}=1.16\times10^{12}\;{\rm M}_\odot$. Logarithmic and NFW estimates are thus 20\% higher and 5\% lower, respectively, than the true mass. These set the upper limit on the accuracy of measuring the total mass of the VL2 halo when assuming these parametric potential forms.

\subsection{Potential evolution in the last 6 Gyr}
\label{sec:evolution}
Dark matter halos experience rapid growth at early times, which slows down after redshift $z\simeq1$. We create streams in Via Lactea II during the slow growth phase, starting from a snapshot 6 Gyr ago (described in the next section). At a final snapshot, streams from this sample are distributed in the inner 150\;kpc. During the 6\;Gyr time span, the VL2 mass has increased $\sim5$\% within 150\;kpc, which is typical for halos in this mass range \citep[private communication]{zemp2013}. In this section we show how the gravitational potential was affected.

\begin{figure}
\begin{center}
\includegraphics[scale=0.5]{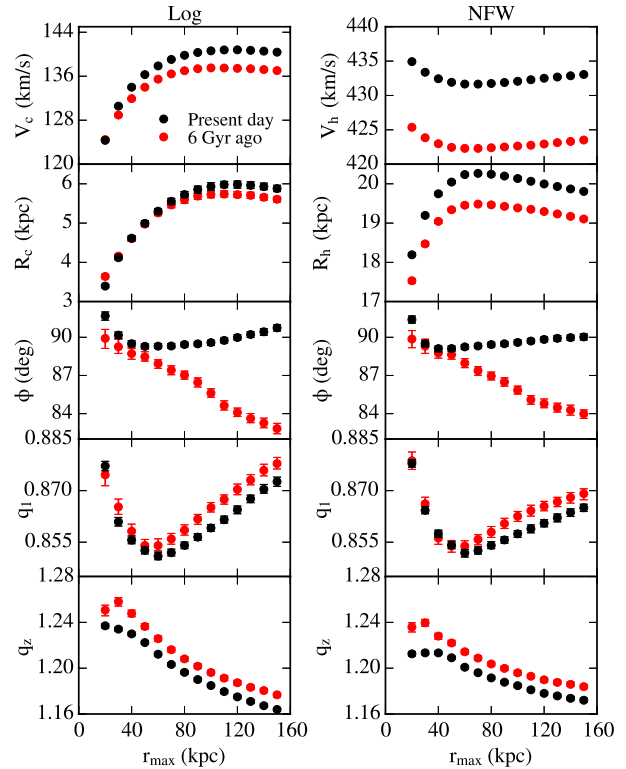}
\caption{The VL2 potential is parametrized with circular velocity ($V_c$, $V_h$), characteristic radius ($R_c$, $R_h$), rotation angle ($\phi$) and axis ratios ($q_1$ and $q_z$), as defined in \S\;\ref{sec:evolution}. We show the best-fit parameters for the triaxial logarithmic (left) and NFW (right) potential to the Via Lactea II simulation, with associated $1\sigma$ uncertainties, measured within a radius $r_{\rm max}$. Even when a complex potential is known exactly, the best-fit parametric form changes with radial extent of the available data. The most significant change in potential parameters from 6 Gyr ago (red) to present day (black) is observed in the normalization of velocity and radius parameters, with minimal changes in the shape parameters. }
\label{potevo}
\end{center}
\end{figure}

We analyze a sample of 100,000 randomly selected simulation particles at the $z=0.69$ snapshot, corresponding to $\sim6$~Gyr lookback time. Analogously to the results of the previous section, we obtain the best-fit parameters for the triaxial logarithmic and NFW potentials as a function of radius, and compare them to the present-day values in Figure~\ref{potevo} (red). The most prominent change is observed in the growth of halo circular velocity and the scale radius. On the other hand, surprisingly little evolution is seen in the shape parameters during this time. 

\section{Simulated streams}
\label{sec:streams}
In this work we focus on cold streams originating from globular clusters. The physics of globular cluster disruption has been extensively studied in direct $N$-body simulations \citep{heggie2003, baumgardt2003, kupper2008a, just2009, kupper2010}, so the formation of their tidal tails is well understood and easier to model than disruption of dwarf galaxies. These simulations show that clusters on a wide range of orbits in a galactic tidal field experience mass loss through evaporation, driven by two-body relaxation. Evaporated stars occupy a range of orbits similar to, though not quite the same as, the progenitor's orbit, thus forming a thin and cold stellar stream \citep{eyre2011}.

The disruption of globular clusters has been studied in $N$-body simulations that directly solve forces between the star particles, but assume a smooth, time-independent galactic potential.
This is at odds with the predictions from cosmological simulations of structure formation -- galactic potentials are clumpy and constantly evolving \citep[e.g.,][]{springel2005}. So far, it has been unfeasible to couple simulations of globular cluster disruption and galaxy formation. The former require high temporal and spatial resolution ($\sim$\;yr, $\sim$\;pc) to correctly capture the details of evaporation, while the latter follow phenomena on much longer ($\sim10$\;Gyr) and larger ($\sim1$\;Mpc) scales. A recent step to bridge this gap follows dynamical evolution of globular clusters using a direct $N$-body simulation, with the tidal forces along their orbits taken from a larger simulation of galaxy formation \citep{renaud2013, rieder2013}. While this approach accurately represents internal globular cluster evolution, the galactic potential, and its influence on the resulting stream, is represented only coarsely by interpolating between the simulation outputs. 

The alternative approach, tailored for studying cold tidal streams in a realistic galaxy potential, is to couple an approximate treatment of the stream formation to a high resolution simulation of galaxy formation. We follow \citet{kupper2012} and create stream models using the {\it streakline} method. Since clusters experience constant mass loss due to two-body interactions and decreasing unbinding energy, we can assume that stars escape the cluster at every simulation time step and trace their paths in the galactic tidal field. The stars escape the globular cluster from a tidal radius, $r_t$, which can be calculated analytically \citep{king1962} for a wide range of galactic potentials and cluster orbits using:
\begin{equation}\label{eq:rt}
r_t = \left(\frac{GM}{\Omega^2-\partial^2\Phi/\partial R^2} \right)^{1/3},
\end{equation}
where $M$ is the cluster's mass, $\Phi$ is the galactic potential, $R$ is the cluster's galactocentric distance, and $\Omega$ is its instantaneous angular velocity. Escaping stars are given velocities to match the cluster's angular velocity, as this choice was shown to reproduce stream morphologies from direct $N$-body simulations \citep{kupper2012}. 

To assess the role of different aspects of the galactic potential in tidal stream evolution, we use a large set of streams whose progenitors are on a variety of orbits. In the rest of this section we describe the two sets of streams: \emph{VL2 streams}, evolved in a live dark matter potential (\S\;\ref{sec:vl2streams}) and \emph{analog streams}, a control set evolved in a parametric potential (\S\;\ref{sec:control}).

\subsection{Via Lactea streams}
\label{sec:vl2streams}

A sample of 12,800 streams was created in a resimulation of the Via Lactea II using the streakline method, starting from particle positions 6~Gyr ago (K\" upper et al., in prep). All streams have progenitors with the same, Palomar~5-like properties; a mass of 20,000$\;\rm M_\odot$ and an effective radius of 20\;pc. Palomar~5 was chosen as an exemplary Milky Way globular cluster that features long tidal tails \citep{odenkirchen2003}. Initially, the clusters were uniformly placed at their orbital apocenters in the $15-150$\;kpc range. Their initial velocities were also uniformly distributed in the $0.25-1\;V_c$ range, where $V_c$ is the local circular velocity. The angular position and the direction of the velocity vector were randomly chosen on a sphere. The clusters' orbits were integrated together with the dark matter particles for 6\;Gyr, with time steps fixed to $dt=1$\;Myr. At each time step, two star particles were released from the cluster's Lagrange points, with the distance from the cluster given by Eq~\ref{eq:rt}, and the gravitational potential evaluated numerically. To create more realistic streams, the streakline method described above was modified to reflect the spread in velocities of stars escaping the cluster \citep{lane2012}. An initial radial velocity offset is randomly drawn from a Maxwell-Boltzmann distribution with dispersion equal to the velocity dispersion in the progenitor \citep{kupper2008b}, which is equal to $\sigma\simeq2$\;km/s for our choice of clusters. Note that, as the stream gets older, its velocity dispersion naturally decreases due to phase-mixing \citep{helmi1999}. The stream stars were treated as massless particles, evolved in a potential created by dark matter particles and globular clusters, but affecting neither. The cluster masses remain constant throughout the simulation.

\begin{figure}
\begin{center}
\includegraphics[scale=0.22]{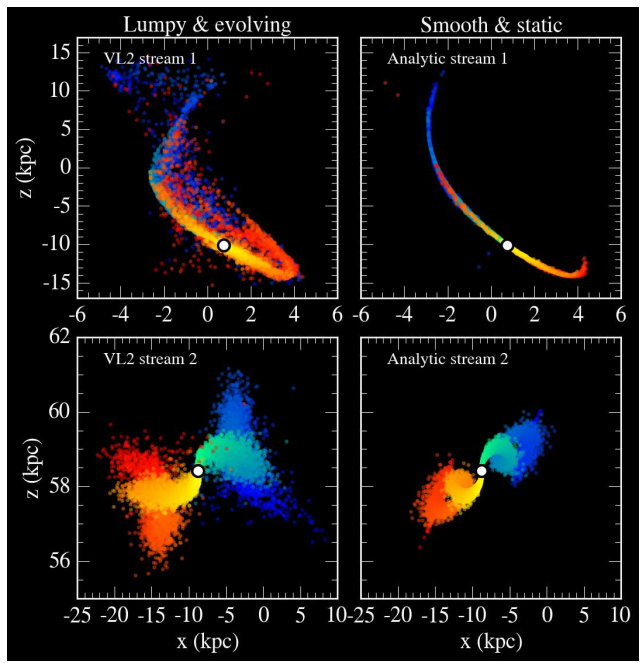}
\caption{Top: Example of a stream close to the pericenter formed in a live dark matter simulation (left) and in an analytic potential (static and smooth, right). Leading tail is shown in warm, trailing tail in cold colors, and the progenitor in white. Darker colors (blue and red) correspond to older parts of the stream, while lighter colors (green and yellow) indicate most recently stripped stars. Bottom: Same as above for a different stream near its apocenter. Analytic streams have similar general shape as the VL2 streams, but are in general narrower.}
\label{fig:streams}
\end{center}
\end{figure}

In this study, we analyze a sub-sample of 256 streams described above, selected to cover the original sample's range in apocentric distance and orbital eccentricity. Two streams from this sample are projected in a $x-z$ plane in the first column of Fig~\ref{fig:streams}. These streams are representative of the sample, chosen to illustrate how streams differ depending on their orbital phases (near pericenter on top, near apocenter on bottom). The stars forming the leading tail are shown in warm colors, and those in the trailing tail are drawn in cold. In both schemes darker colors represent older parts of the stream. The progenitor is marked with a white circle. The top stream is near its pericenter, and thus very extended. The bottom stream, close to its apocenter, spans a much smaller distance range. The density along the stream is non-uniform in both cases, which is due to the combined effects of epicyclic motion of stream stars with respect to the progenitor's orbit \citep[e.g.,][]{kupper2008a} and encounters of stream stars with the dark matter subhalos \citep[e.g.,][]{carlberg2013}. The former effect is a consequence of the cluster disruption mechanism, so it is also present in smooth potentials. The latter effect is inherent to clumpy potentials. 

\subsection{Analytic stream sample}
\label{sec:control}
To illustrate the impact of a live dark matter potential on the stream morphology, we contrast our sample to one evolved in a static and smooth potential. This control sample consists of 256 streams, with progenitors currently at the same position and orbital phase as in the original sample, but evolved in an analytic potential. For the analytic potential, we choose the best-fit logarithmic and NFW form to the global, present day VL2 potential (black points at 150~kpc in Fig~\ref{potevo}). The progenitors' orbits were traced back for 6\;Gyr in the chosen potential. Once the initial position and velocity of the progenitor had been determined, the streakline method was used to create streams analog to the VL2 sample, with the same progenitor properties as described above. 

In the right column of Fig~\ref{fig:streams} we show smooth streams, analog to the realistically evolved ones in the left column. The range spanned by the streams and their general shape are remarkably similar between the samples. The major difference is the streams' widths, with streams from the VL2 simulation being wider due to the combined effects of potential evolution and heating by the dark matter subhalos. 

\section{Stream modeling}
\label{sec:modeling}
Our goal is to use streams to infer the galactic potential. In this section, we present a method for measuring the potential using the ``forward modeling'' of tidal debris from globular clusters. Our \ff method makes models of a stellar stream in a test potential and compares the resulting phase-space distributions to the observed stream. The probability of an observational data set being generated from these distributions is used as the likelihood in a Markov Chain Monte Carlo (MCMC) algorithm to efficiently search for the underlying probability distributions of the potential parameters.
We are primarily interested in the accuracy of the realistic potential recovery when assuming a simplified form.
Therefore, we apply the \ff method to our stream samples in the absence of observational errors. However, the method can be naturally extended to include those errors. A similar method was used by \citet{varghese2011} and \citet{gibbons2014} to model debris originating from more massive progenitors, and within the action-angle framework by \citet{bovy2014} and \citet{sanders2014}. Below we describe the method in more detail (\S\;\ref{sec:ff}), and test it on a $N$-body simulation of a Palomar 5-like stream (\S\;\ref{sec:pal5mock}).

\subsection{The \ff method}
\label{sec:ff}
The first step in the \ff method for constraining the potential is to create a model of the stream under consideration. Assuming the current position and velocity of the progenitor are known accurately, we first track the progenitor back along its orbit in a test potential for 6\;Gyr, the length of time the streams have been forming in our sample. Then, the streakline method is used to create a realistic stream model. To efficiently model the whole stream sample, we used a fast leapfrog integrator with constant time steps of 1\;Myr, which corresponds to the time step used in the VL2 resimulation.

For each model stream created in a test potential, we assess the likelihood of this particular set of parameters, by comparing the model stream to the observed one. In this work, the ``observed'' streams are the streams simulated in the VL2 (\S\;\ref{sec:vl2streams}) or their analytically evolved analogs (\S\;\ref{sec:control}). Naturally, the same method can be applied to streams observed in the Galactic halo and beyond. We assume there are $N$ member stars observed in each stream, with accurate positions and velocities. This we call our data. As a shorthand, we denote this 6D phase space $X$, and data points in it $X_n$. Within the framework developed by \citet{hogg2010} and \citet{hogg2012}, the probability of a model point $x_k$ generating a data point $X_n$ is then:
\begin{equation}
p(X_n|x_k,\theta,I)=\mathcal{N}(X_n|x_k, \sigma_n^2+\Sigma_k^2) ,
\label{eq:p}
\end{equation}
where $I$ represents any prior information about the problem, $\mathcal{N}(x|m,V)$ is a 6-dimensional Gaussian distribution for $x$ given mean $m$ and variance tensor $V$, $\sigma_n^2$ is the observational variance tensor, while $\Sigma_k^2$ is a smoothing variance tensor that effectively turns the finite set of model points into a smooth density function in phase space, $\theta$ are the potential parameters. If the progenitor's orbit and phase information are unknown, they can be included in $\theta$ as initial-condition parameters. We treat the smoothing between the model points, $\Sigma_k$, as a hyper-parameter in our modeling process, and marginalize over it when reporting results. For simplicity, we assume that the observations are perfect and there are no observational errors. However, this formalism allows for inclusion of realistic errors by simply setting $\sigma_n\neq0$.

The likelihood of the data point $X_n$ being represented by the whole of the model stream, is calculated by marginalizing out the model points $x_k$ in Eq~\ref{eq:p}:
\begin{equation}
p(X_n|\theta,I)=\sum_{k=1}^K P_k p(X_n|k,\theta,I) ,
\label{eq:pi}
\end{equation}
where we use $K$ model points, equally weighted such that $P_k=1/K$. The precision of the recovered potential parameters increases with the number of data and model points used, but so does the modeling run time. In order for the likelihood function to be smooth and easily sampled, the number of model points should be several times larger than the number of data points. An empirical study of the combination of these effects on convergence, precision and run time led to the chosen setup with $N=100$ data points and $K=500$ model points. This somewhat arbitrary choice can be easily modified to optimally utilize available computational resources.

Finally, we assume that the data points are independent. The likelihood of all of the data points being drawn from the model distribution is, in this case, a product of their individual likelihoods given by Eq~\ref{eq:pi}:
\begin{equation}
P(\{X_n\}|\theta,I)=\prod_{n=1}^N p(X_n|\theta,I) .
\label{eq:ptot}
\end{equation}
The natural logarithm of likelihood defined by Eq~\ref{eq:ptot} is used as an input for the affine invariant MCMC algorithm \citep[\textit{emcee}, ][]{mc}. The parameter space of $\theta$ is explored with 150 walkers. Convergence is obtained after a short burn-in phase of 100 steps. The chains are then restarted around the best-fit value, and evolved for another 1000 steps. The convergence was ensured by unchanging values of mean and standard deviation among the walkers in the third and fourth quartile. We report the median values and $1\sigma$ spreads of the resulting parameter distributions in the following sections.

\subsection{$N$-body test of the \ff method}
\label{sec:pal5mock}
Since both our stream generation and potential recovery are based on the streakline method, it is crucial to establish whether our method correctly recovers gravitational potentials using a different stream model. To demonstrate that the \ff method works for non-streakline streams, we test it on a $N$-body simulation of a Palomar 5-like stream before applying it to the stream data sets described in \S~\ref{sec:streams}. 

The test stream was evolved for 6\;Gyr in an analytic Milky Way-like potential using a modified version of the direct $N$-body code NBODY6 \citep{aarseth2003}. The potential includes contributions from a disk, a bulge and a flattened, axisymmetric NFW halo. In modeling this stream, we fix the baryonic contribution to the gravitational potential, and only fit the dark matter halo, which we represent as a three-parameter NFW potential (Eq~\ref{nfw} with fixed orientation angle $\phi=90^\circ$ and $y$-axis flattening $q_1=1$). The best-fit model, obtained through \ff modeling described above, accurately recovers the underlying potential. The recovered values of the potential parameters are listed in Table~\ref{tb:fftest} and compared to their respective true values. The halo circular velocity and flattening are recovered very accurately (less than 1\% error), while the errors in the scale radius are on the order of $\lesssim5\%$. This discrepancy in the recovery of global parameters can be explained by the progenitor's orbit being inside the scale radius of the halo potential. Since the stream is not probing the region where the potential power-law slope breaks, it is less sensitive to the value of the scale radius. This test shows that the \ff method can be used to reliably recover potential parameters, but also indicates that some streams are less well suited for recovering halo parameters than others.

\begin{table}
\caption{\ff potential recovery using a $N$-body stream}
\label{tb:fftest}
\begin{center}
\footnotesize{
\begin{tabular}{l c c c}
\hline\hline
Parameter\footnotemark[1] & \shortstack{True \\ value} & \shortstack{Recovered \\ value} & \shortstack{Ratio\footnotemark[2]}\\
\hline
Circular velocity (km/s) &   417 &   418$^{+   26}_{-   17}$ &  1.00 \\ 
Flattening $q_z$ &  0.94 &  0.94$^{+ 0.01}_{- 0.01}$ &  1.00 \\ 
Scale radius $r_h$ (kpc) & 36.54 &    38$^{+    7}_{-    4}$ &  1.04 \\ 

\hline
\end{tabular}
\footnotetext[1]{Parameters of a NFW halo (as defined by Eq~\ref{nfw}, assuming a fixed orientation angle $\phi=90^\circ$ and $y$-axis flattening $q_1=1$).}
\footnotetext[2]{Recovered / True}
}
\end{center}
\end{table}

\section{Results}
\label{sec:results}
In this section we present results of applying the \ff method for potential recovery on our stream samples. We start by illustrating the fitting procedure and posterior parameter distributions obtained using a single VL2 stream in \S\;\ref{sec:res:1}. The same procedure is then repeated for all of the streams, and the summary results are presented in \S\;\ref{sec:res:an} for analytic streams, and \S\;\ref{sec:res:vl2} for the VL2 stream sample.

\subsection{Potential recovery example using a single stream}
\label{sec:res:1}
\begin{figure*}
\begin{center}
\includegraphics[scale=0.51]{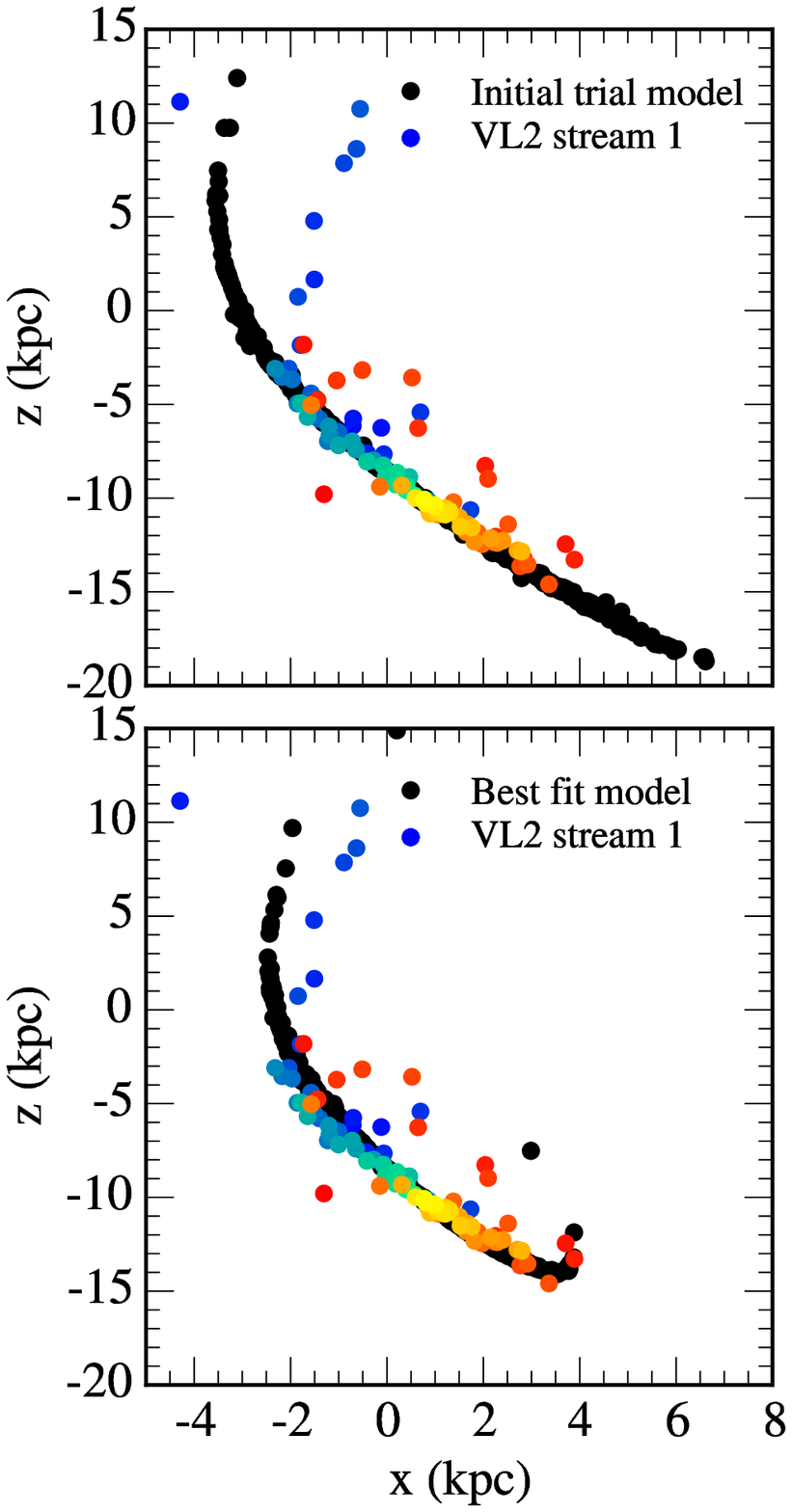}
\includegraphics[scale=0.19]{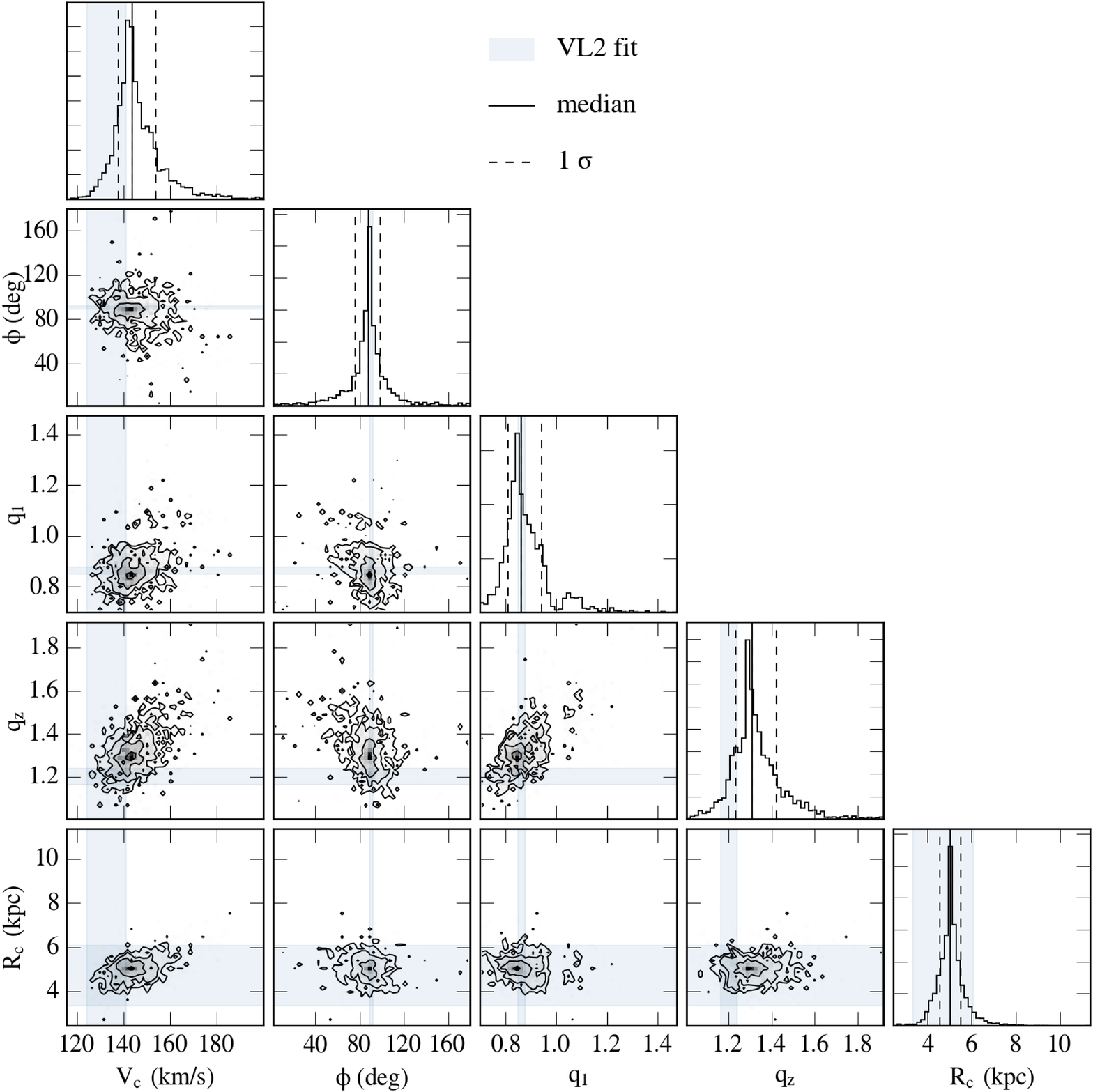}
\caption{Modeling results for VL2 stream 1 (top of Figure~\ref{fig:streams}). Top left: Initially, the trial potential produces an unlikely stream model (black) for the observed stream (colors). Bottom left: The best-fit model is a good match of the observed stream. Right: Posterior parameter distributions in one (histograms) and two dimensions (gray contours). Range of values recovered by direct fits to the VL2 simulation is shown in solid blue, while black solid and dashed lines represent the median and 1$\sigma$ intervals of the parameter distributions. For this stream, all of the parameter estimates are in $1\sigma$ agreement with the expectations from fits to the VL2 data.}
\label{fig:diags}
\end{center}
\end{figure*}

We first apply the \ff method on a single VL2 stream, recover the underlying potential assuming a logarithmic form, and show detailed constraints on the potential parameters. To illustrate how streams can discriminate between the different logarithmic potentials, we show a VL2 stream observed in the $x-z$ plane in the left panel of Figure~\ref{fig:diags}. On top, we compare the stream (colors) to a model (black) created in a trial potential. This model is not a good representation of the observed VL2 stream because it extends past the observed range. More precisely, many of the model stars are far away from the observed stars, so this model has a low likelihood (calculated using Eq~\ref{eq:pi}). The MCMC sampler searches the parameter space for a potential that maximizes the likelihood, and such a best-fit model is shown on the bottom of Figure~\ref{fig:diags}. It matches the observed stream well.

Once the best-fitting potential parameters are located, the MCMC sampler explores the surrounding parameter space to recover the probability distributions of parameters. On the right of Figure~\ref{fig:diags} we show posterior probability distributions of potential parameters obtained by modeling the VL2 stream shown on the left. Panels on the diagonal show the one dimensional parameter distributions as histograms, overplotted with the median (solid vertical line) and 68\% uncertainty limits (dashed vertical lines). Subsequently, we quote the median as our best estimate of a parameter.

The range of parameters expected from fitting the present day VL2 potential directly (full radial range in Figure~\ref{potevo}, \S\;\ref{sec:fits}) is shown as shaded blue regions in Figure~\ref{fig:diags}. For this stream, all of the recovered parameters are within $1\sigma$ of the expected values. Additional information about the covariance between the potential parameters is presented on 2D probability distributions (lower left panels). Color of the contours corresponds to the probability density, with darker gray corresponding to more probable regions of the parameter space. All the walkers have converged to a single solution and there are no indications of multimodality in the parameter distributions. There is some correlation between circular velocity and core radius, while other parameters are mostly uncorrelated for this stream. In the next sections we present results for the whole sample of analytic and VL2 streams. 

\subsection{Streams evolved in analytic potentials}
\label{sec:res:an}
We proceed to apply the \ff potential recovery method on our analytic stream samples, where the potential is smooth and static, and its form is known. One set of streams was created in a logarithmic potential, other in a NFW potential (\S\ref{sec:control}). Each stream from these samples was modeled individually in the appropriate potential as described above, and here we report the recovered potential parameters for the whole sample.

\begin{figure*}
\begin{center}
\includegraphics[scale=0.26]{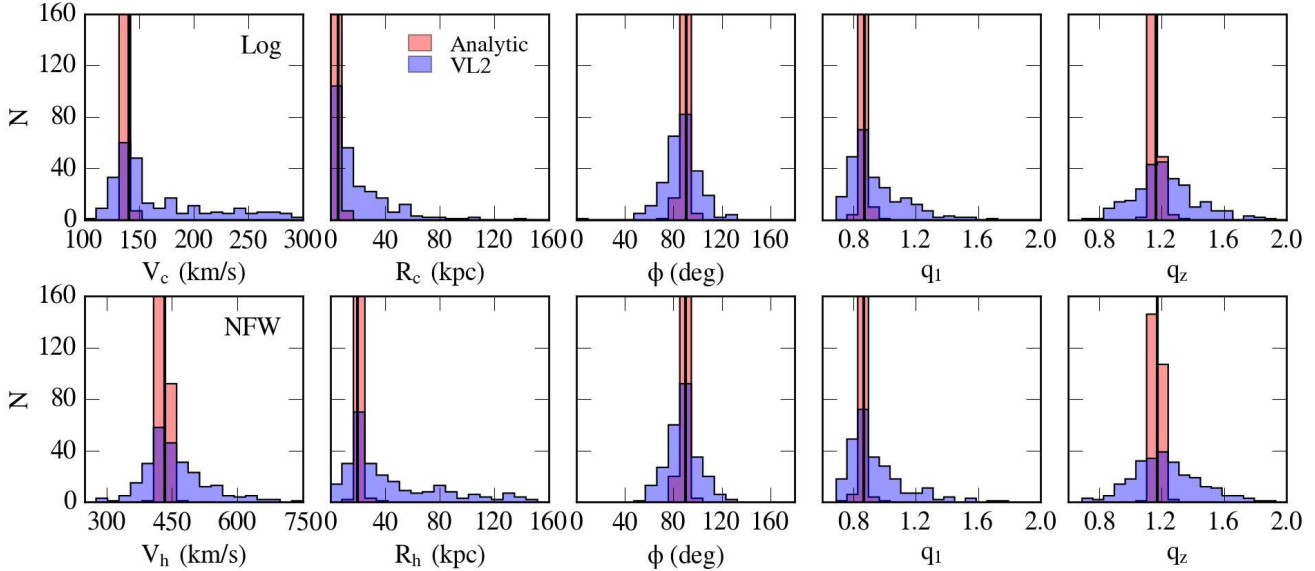}
\caption{Distributions of best-fit parameter values across a sample of 256 streams, logarithmic on top and NFW on bottom. Analytic streams are shown in red, while the VL2 streams are blue. The true parameter values are marked with vertical black lines. The distributions of both the analytic and the VL2 streams peak at the true values, with VL2 streams having larger spread around the mean.}
\label{fig:pdfs}
\end{center}
\end{figure*}

The distribution of median potential parameters recovered using all analytic streams individually is shown with red histograms in Figure~\ref{fig:pdfs}. Streams created in a logarithmic potential are on the top, while those formed in a NFW potential are on the bottom. The input parameter values are marked on each panel with vertical solid lines. In both potentials, most of the streams recover the true values. The distributions' modes coincide with the true values, while the medians are within 3\% for all of the parameters. The largest discrepancies are seen in core radius for the logarithmic potential, and in scale radius for the NFW potential. Fractional errors in other parameters are smaller than 10\% for more than 98\% of the streams in both potentials.

We conclude that true potential parameters are accurately recovered using the \ff method on streams evolved in an analytic potential. In addition, the method provides uncertainties for the recovered parameters. Typical 68\% uncertainty range in a parameter's distribution function, which we also call the $1\sigma$ error, is small, on the order of $\lesssim10\%$. These errors are only slightly overestimated, since $\sim90\%$ of streams in our sample recover true parameter values on the $1\sigma$ level. Combined, these results reveal no biases inherent to the \ff potential recovery method. Its estimates of potential parameters using analytic streams on a variety of orbits are both accurate and precise.

\subsection{Streams evolved in the VL2 potential}
\label{sec:res:vl2}
In this section we present potential recovery results for streams evolved in a clumpy and evolving VL2 potential, but modeled in smooth and static potentials. The distributions of potential parameters are overplotted as blue histograms in Figure~\ref{fig:pdfs} (logarithmic potential on top, NFW on bottom). Parameter values marked with vertical solid lines were initially obtained by fitting the logarithmic/NFW forms directly to the present day VL2 potential inside 150~kpc.

Similar to the analytic streams, modes of the parameter distributions for the VL2 streams are centered on the true values of the global, present-day VL2 potential. In this case, however, the distributions are not as symmetric, so their medians are within $\sim$10\% of the true values. The largest deviation is seen in scale radius for the NFW profile, driven by overestimates coming from streams on orbits far beyond the true scale radius, and thus less sensitive to this parameter. In general, the distributions are wide and only $40-60$\% of the VL2 streams recover parameters within 10\% of the true values (compared to 90\% of analytic streams). Correspondingly, the uncertainties in recovered parameters are also larger for VL2 streams, on average 15\%. 
However, we find our errors are an accurate assessment of measurement uncertainties, as $\sim$50\% of the streams recover true parameters within 1$\sigma$. Some streams are still individually biased, but collectively they recover global properties of the present day VL2 halo (blue histograms on Figure~\ref{fig:pdfs}), even when we assume a smooth and static analytic potential in their modeling. 

This collective behavior makes it reasonable to assume that potential constraints obtained by modeling multiple streams simultaneously will narrow down the parameter distributions. Indeed, \citet{deg2014} have shown that potential recovery with two streams improves on the individual results for the halo shape parameters. The Milky Way streams have so far been modeled only individually, yielding different estimates for the dark matter halo shape \citep[e.g., ][]{law2010,koposov2010}. Joint stream constraints will be crucial in measuring such a complex potential.

\subsection{Mass estimates using streams}
Mass is the most basic property of dark matter halos. We have so far demonstrated how tidal streams can constrain the gravitational potential, and in the following we will use these constraints to determine the mass of a halo. For brevity, we focus on mass within 150\;kpc, but similar conclusions can be reached for any of the global halo parameters. To get the halo mass, we follow the procedure outlined in \S\;\ref{sec:fits}; first calculating the density field from the potential using the Poisson equation, and then integrating it out to 150\;kpc. This limit was chosen to match the radial extent of our stream sample. 

\begin{figure*}
\begin{center}
\includegraphics[scale=0.245]{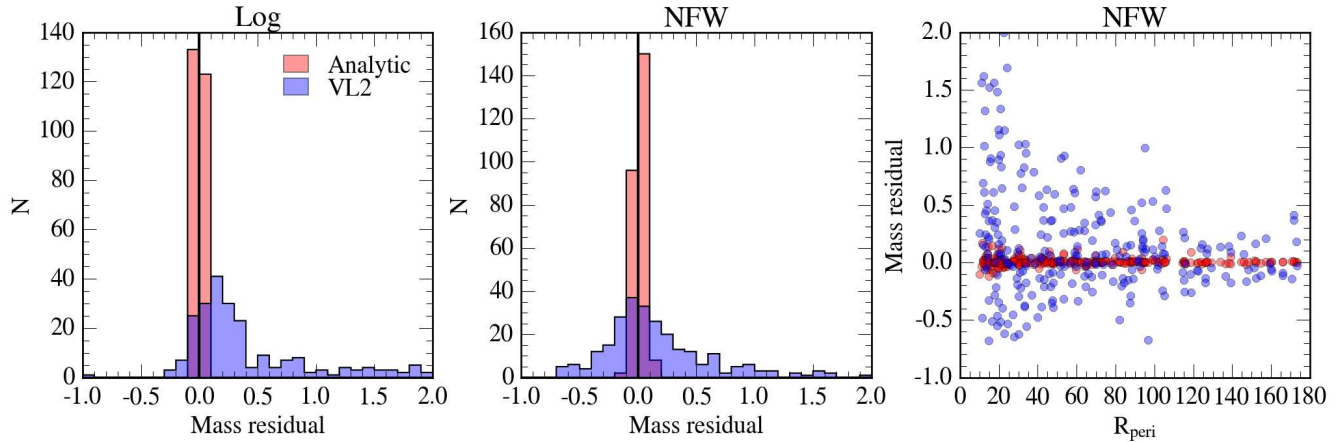}
\caption{Distributions of residuals in mass within 150~kpc when assuming the logarithmic (left) and NFW potentials (middle). Analytic streams (red) recover the true mass very accurately in both potential forms, and their residuals are centered on zero (black vertical line). VL2 streams (blue) are biased towards more massive halos in the logarithmic form, and lower mass halos in the NFW potential. Large residuals are more likely for streams with smaller pericentric distances in both potentials; to reduce clutter we present only NFW results (right). }
\label{fig:mass}
\end{center}
\end{figure*}

Analytic streams accurately measure the halo mass. Their mass residuals, defined as the difference between the stream estimate of mass and the actual value, normalized to the actual mass, are shown in red in Figure~\ref{fig:mass} (left panel for logarithmic potential, middle for NFW). The distributions have the mode at 0\% error in both logarithmic and NFW potentials. These distributions are also very narrow, with 100\% of the logarithmic and 96\% of the NFW streams having less than 10\% error in mass. Accurate mass determination with analytic streams is not surprising, given their accurate recovery of individual potential parameters. However, it is encouraging that the mass, a calculated global quantity, is recovered with the same precision as the parameters being solved for by stream modeling.

Residuals in mass calculated using the VL2 streams are shown in blue histograms in Figure~\ref{fig:mass}. These distributions are wider than the corresponding analytic ones, as has already been seen in distributions of potential parameters. Unlike the analytic stream sample, VL2 mass estimates are biased to higher masses when assuming a logarithmic potential (mode at 20\%), and to slightly lower masses when assuming a NFW potential (mode at $-5$\%). These values are comparable to discrepancies in mass predicted by logarithmic and NFW fits to the VL2 potential directly (18\% and $-6$\%, respectively). Therefore, assuming an analytic potential limits dark matter halo mass measurement to an accuracy of 5 to 20\%, depending on the choice of analytic parametrization.

Apart from the distribution median displacement in the VL2 sample, there is also a tail of high mass outliers. These correspond to streams with high recovered circular velocities and scale radii. We explore the origin of these outliers by showing mass residuals for VL2 streams, modeled in a NFW potential, as a function of closest pericentric passage of their progenitors in the right panel of Figure~\ref{fig:mass}. All of the high mass outliers originate from streams with small pericentric radii, while the streams that never enter the inner $\sim70$~kpc have mass residuals smaller than 50\%, motivating the search for streams at large galactocentric radii. VL2 streams modeled in a logarithmic potential show a similar trend, so we omit them from this panel to reduce clutter. The fraction of streams with extremely large mass estimates is small; most streams in the inner region of VL2 produce errors in mass smaller than 30\%. This is encouraging for measuring the mass of the Milky Way, since all of the Galactic cold streams have so far been discovered within 30 kpc.

\section{Discussion}
\label{sec:discussion}

\subsection{Comparison of VL2 and analytic streams}
The presence of substructure in realistic dark matter halos is the main qualitative difference between the VL2 and analytic stream models. Accretion of subhalos over cosmic time causes the potential of the host halo to evolve, and their subsequent motion within the halo adds to a time dependent, clumpy potential, which is only approximately described by analytic forms. In potential recovery using streams, the effects of (1) substructure, (2) potential evolution, and (3) inadequate potential form in stream modeling result in wider parameter distributions for the VL2 sample than the analytic case. Below we look into each separately.

\emph{Substructure:} Although they contribute about 10\% of the dark matter in Milky Way-mass halos, subhalos occupy only a small fraction of the total halo volume. To assess their impact on the streams, we count the total number of their encounters in the 6\;Gyr of evolution. Assuming that an encounter happens when a subhalo passes within $\sim1.5\;$kpc of a stream, the streams in our sample have had on average 2.4 encounters with dark subhalos more massive than $10^5\;\rm M_\odot$. Although no clear trend between the error in mass estimate and the number of stream--subhalo encounters is observed, streams that experienced encounters with massive subhalos ($M_{\rm sub}\gtrsim10^7\rm M_\odot$) are more likely to result in overestimates of halo mass. These massive encounters can disrupt parts of the stream, making the coherent part appear shorter. In general, if all the other parameters are kept fixed, globular clusters evolved in a more massive halo have smaller tidal radii, and consequently develop shorter tidal tails. This explains the tendency to overestimate halo mass using steams which have encountered massive subhalos. The stream length also depends on its age, which is unknown in a realistic case, so this effect of substructure might be even less apparent. In summary, while encounters with subhalos can significantly alter the stream and affect subsequent potential recovery, this only happens for a small subset of streams ($\lesssim$15\% have had more than five encounters). Hence, halo substructure only marginally contributes to the spread in recovered potential parameters across the VL2 stream sample. 

\emph{Evolution:} The continuing accretion of substructure on dark matter halos produces potential evolution that is rarely modeled when attempting to recover gravitational potentials (however, see \citealt{buist2014} for a recent formulation that takes into account growth of the scale mass and radius). As shown in \S\;\ref{sec:evolution}, the shape of the VL2 halo has experienced little change in the last 6\;Gyr, but both mass and scale radius have increased by $5-10\%$. Although all of the streams have been exposed to the growth of the host dark matter halo, most recover present day values of scale mass and radius. Similarly, \citet{penarrubia2006} found that streams adapt to adiabatic potential changes, so in the case of smooth and slow potential evolution the present-day information on streams only yields constraints on the present-day potential. However, the change of the potential parameters in the last 6\;Gyr makes up for only $\sim20\%$ of the spread in potential recovery results using the VL2 stream sample. The spread in obtained parameters is only in part due to potential evolution.

\emph{Potential form:} Another aspect of simulated dark matter halos that is poorly captured by analytic models are the radial changes in flattenings and orientation angles \citep[e.g.,][]{allgood2006}. When fitting such a halo with a potential that only has global parameters for flattening and orientation, the output depends on the radial extent of available data, as shown in \S\;\ref{sec:fits}. The radial variations in the VL2 shape parameters are on the order of $20-50\%$ of the FWHM spread in recovered shape parameters across our stream samples. Thus, inadequate potential form used in modeling streams can also only partly explain the observed spread in recovered parameters.

Each of these effects contributes to the spread in recovered potential parameters. We estimate that the role of direct encounters with subhalos is minor, but potential evolution and a simplistic potential form individually contribute to the spread on a $20\%$ level. Combined, radial variations in the VL2 shape parameters in the last 6\;Gyr make up for more than $\sim70\%$ of spread in recovered parameters. The spread in recovered distributions of potential parameters appears to be driven by the potential evolution and by streams probing different radii in the host halo. Studying these effects in dedicated simulations with only substructure or only smooth halo growth will provide a clearer picture on the impact each has on potential recovery. 

The combined effects of potential clumpiness, evolution and incorrect form result in biased total mass estimates when replacing the realistic potential with an analytic approximation. The measurement is equally biased when constraining the potential by applying the \ff method on our VL2 stream sample, or directly fitting analytic potentials to the VL2 particle data.
We note that many different classes of potential recovery methods using tidal streams have been proposed recently \citep[e.g.,][]{sanders2013b,apw2013,deg2014,bovy2014,sanders2014}. 
While the details of potential recovery using these other methods may be differently affected by the assumption of a smooth and static potential, our conclusions regarding the mass bias should remain valid for all potential recovery methods based on this assumption.

\subsection{Implications for known Milky Way streams}
\label{sec:mw}
In this section we discuss the error in mass estimates induced by the assumption of a smooth and static potential when modeling real Milky Way streams. While Galactic streams are influenced by both dark matter and baryons, streams in our sample were evolved in dark matter-only potentials. Hence, we can only assess the errors in the dark matter component, which will make up only part of the error budget in the estimates of the total galaxy mass.

The most prominent streams originating from globular clusters in the Milky Way are the Palomar~5 tidal tails and GD-1. Based on kinematic observations of the cluster and the position of its tails, \citet{odenkirchen2001} estimated the Pal~5 pericenter and apocenter distances of $7\;$kpc and $19\;$kpc, respectively. Similarly, \citet{koposov2010} used 6D information on the GD-1 stream to obtain the pericenter and apocenter of $14\;$kpc and $26\;$kpc, respectively. These values are uncertain due to observational errors and dependent on the choice of the Galactic potential. We constructed a conservative sample of Pal~5 and GD-1 analogs from our VL2 sample by requiring that their peri- and apocenters are matched within 5\;kpc. The median mass residual of these samples is small ($\lesssim5$\%), but the dispersion within the sample is $\sim$50\%. In other words, while these streams collectively provide an accurate measurement of the host halo mass, results for the individual streams can be heavily biased. 

Total halo mass measurements using tidal streams evolved in a complex, dark matter-only potential are more accurate when distant ($d>70\;$kpc) streams are employed (see Figure~\ref{fig:mass}). 
In a realistic Galactic potential, dark matter halo potential recovery using streams with small pericenters is further impeded by the disk. Although the effect of dark matter subhalos on streams at small galactocentric radii ($d<20\;$kpc) is expected to be lower due to their depletion by the disk \citep{donghia2010}, the potential is dominated by the disk, so streams at those distances might not be very sensitive to global halo shape \citep{koposov2010}. Consequently, distant streams are also expected to provide better handle on the global halo potential in a galactic potential. While there are no distant cold streams known at the moment, the outer Milky Way halo will be mapped down to faint magnitudes in the LSST era \citep{ivezic2008}. Any streams discovered in upcoming deep surveys will provide valuable anchors for the total mass, and combined with the nearby streams, they will improve constraints of the entire Galactic potential. 

\section{Summary}
\label{sec:summary}
In this study, we revisited gravitational potential recovery using tidal streams. In particular, we tested the performance of analytic potentials in representing a realistic dark matter halo. In doing so, we developed a novel method to forward model streams originating from globular clusters, with realistic stream morphologies and stream--orbit offsets. We applied this method to two samples of streams, one evolved in an analytic and the other in a $N$-body potential. We modeled these streams in a smooth and static potential, and produced estimates of potential parameters and total mass within 150\;kpc. 

The true potential parameters are accurately recovered by our analytic stream samples. In a realistic, $N$-body halo, these parameters change with time and radius. While the streams evolved in such a halo collectively recover the present day, global values, the distributions of individual estimates are wide. The width of these distributions is comparable to the parameter variations in the radial range probed by the streams during their evolution. We estimate that the impact of substructure on this widening is marginal, due to scarcity of streams that have been significantly influenced by subhalos.

The global halo mass is also accurately measured using streams in the analytic case, but biased in a realistic halo. This bias stems from the analytic potential only approximating the true potential form, and is independent of the potential recovery method. The amount of bias depends on the choice of such a potential, and is lower for the NFW ($\sim5\%$) than for the logarithmic potential ($\sim20\%$). In addition to the overall bias, streams passing through small galactocentric distances are shown to have larger errors in mass estimates. Streams beyond 70\;kpc typically measure the total mass with 30\% accuracy.

We are entering an exciting age for Galactic studies. Near-future missions like Gaia, DES and LSST will accurately map a large volume of the Milky Way, providing a detailed look into streams populating a large range of galactocentric distances. Using the methods outlined in this paper, we will be ready to model these streams and accurately pin down the Milky Way mass. This will have profound implications for placing the Milky Way in a cosmological context.

\vspace{0.5cm}
\emph{Acknowledgments:} The authors wish to thank Julianne Dalcanton for the initial project idea, Marcel Zemp for measuring typical growth of Milky Way-massed halos, Adrian Price-Whelan for help with running \emph{emcee}, Dan Foreman-Mackey for discussion of errors, Erik Tollerud for discussion of MCMC, Duncan Campbell, Kareem El-Badry, and Andrew Hearin for comments on the manuscript. We also thank Amina Helmi, Jorge Penarrubia and Andreea Font for insightful conversations, and Gaia challenge workshop for hospitality. This work has made use of the SuperMUC and zBox supercomputers at LRZ Garching, Germany and University of Z\" urich, Switzerland. 
AB, AHWK and MG acknowledge support from HST-GO-12957.04-A provided by NASA through a grant from STScI, which is operated by AURA, Inc., under NASA contract NAS 5-26555.
AHWK would like to acknowledge support through the DFG Research Fellowship KU 3109/1-1, and support from NASA through Hubble Fellowship grant HST-HF-51323.01-A awarded by the Space Telescope Science Institute, which is operated by the Association of Universities for Research in Astronomy, Inc., for NASA, under contract NAS 5-26555.
JD is supported by the Swiss National Foundation. KVJ's contributions were supported by NSF grant AST-1312196. DWH was partially supported by NASA (grant NNX12AI50G), the NSF (grant IIS-1124794), and the Moore--Sloan Data Science Environment at NYU.

\bibliographystyle{apj}
\bibliography{apj-jour,live_modeling}

\end{document}